\documentclass[journal=jpcbfk,manuscript=article]{achemso}

\usepackage{amsmath}
\usepackage{amssymb}
\usepackage{graphicx}
\usepackage{bm}
\usepackage{color}
\usepackage{hyperref}

\newcommand{\kB}{k_{{\tiny B}}}

\newcommand{\Del}{\Delta}

\newcommand{\eps}{\epsilon}
\newcommand{\be}{\begin{equation}}
\newcommand{\ee}{\end{equation}}

\SectionNumbersOn

\author{Eric A Mills}
\author{Steven S Plotkin}
\email{steve@phas.ubc.ca}
\phone{1 604-822-8813}
\fax{1 604-822-5324}
\affiliation{Department of Physics \& Astronomy, University of British Columbia, Vancouver, British Columbia V6T1Z4 Canada}

\title{Density Functional Theory for Protein Transfer Free Energy}

\begin{document}

\abstract{We cast the problem of protein transfer free energy within
  the formalism of density functional theory (DFT), treating the protein as
  a source of external potential that acts upon the solvent. 
Solvent excluded volume, solvent-accessible surface area, and
temperature-dependence of the transfer free energy all emerge
naturally within this formalism, and may be compared with simplified
``back of the envelope'' models, which are also developed here. 
Depletion contributions to osmolyte induced stability range from
$5$-$10\kB T$ for typical protein lengths. 
The general DFT transfer theory developed here may be
  simplified to reproduce a Langmuir isotherm condensation
mechanism on the protein surface in the limits of short-ranged
interactions, and dilute solute. 
Extending the equation of state to higher solute densities results
  in non-monotonic behavior of the free energy driving protein or
  polymer collapse.
Effective interaction potentials between protein backbone or
sidechains and TMAO are obtained, assuming a simple backbone/sidechain
2-bead model for the protein with an effective 6-12 potential with the
osmolyte. 
The transfer free energy $\delta g$ shows significant entropy: 
$d(\delta g)/dT \approx 20 \kB$ for a 100 residue protein. 
The application of DFT to effective solvent forces for use in
implicit-solvent molecular dynamics is also developed. 
The simplest DFT expressions for implicit-solvent forces contain
  both depletion interactions and an ``impeded-solvation'' repulsive force at
  larger distances. \\

}

\section{Introduction}
\label{sec:intro}

Proteins fold and function in the crowded environment of the
cell. Cytosolic proteins must negotiate a complex milieu which in
many ways is significantly different than the environment in the test tube:
roughly 15\% of water molecules are motionally restricted by protein and
membrane surfaces~\cite{PerssonE08}; the surrounding solvent is
enriched in ions such as Potassium but depleted in Sodium and
Chlorine; osmoprotectants such as trehalose and various amino acids
are present in significant concentration; numerous membrane surfaces
such as the nucleus, ER, and Golgi impose charged substrates for
protein interaction; macromolecular agents such as the microtubules,
actin, ribosomes, soluble proteins and RNA occupy roughly 30\%
($\approx 300 \mbox{g/}\ell$) of the
cellular volume, and modulate stability \cite{YanceyPH82}, aggregation
propensity \cite{EllisRJ06}, and dissociation constants~\cite{ZhouHX08arb,MintonAP01}.

Non-cytosolic proteins also fold in environments distinct from the test
tube as well as the cytosol, particularly with respect to ionic and redox conditions as well as the
chaperone complement. Proteins destined for the plasma membrane or
extracellular matrix are trafficked by the secratory pathway through
the ER and Golgi~\cite{Ellgaard2003}. The environments in the ER and
cytosol are sufficiently different that the conditions for protein folding
are generally mutually exclusive between the two milieu. Folding
generally occurs in the lumen of the ER, while function occurs either
on the plasma membrane or in the
extracellular matrix, which is itself densely occupied by highly
charged glycosaminoglycans such as hyaluronan and heparin sulfate---large molecules that may facilitate cellular migration and regulate
secreted protein activity. Fibrous proteins such as collagen and
fibronectin also occupy the extracellular space, and provide
structural rigidity while allowing rapid diffusion of nutrients and
signalling metabolites between constituent cells. 

The above examples demonstrate the need to correctly account for
  the effect of the cell environment on protein folding, stability,
  and function. Accurately accounting for
  the effects of the cell environment presents a challenge however to both experimental and
computational studies. Experimentally, most of what is known about
protein folding and stability has resulted from {\it in vitro} studies
at dilute concentrations, 
and many questions remain as to how well such
results apply to a realistic cell environment. 
Computationally, including explicit solvent along with a
realistic concentration of osmolytes in a box 
of sufficient size to implement periodic boundary conditions outside
the range of an electrostatic cutoff
typically
increases the number of particles in
the simulation by a factor on the order of ten or more~\cite{CanchiDR2010}. While this can be done for small proteins
such as Trp-cage \cite{CanchiDR2010}, investigating larger
proteins generally requires coarse-grained models to keep the
computational resources required reasonable~\cite{LinhanantaA11}. 

Computational studies of crowding on isolated monomeric minimal
$\beta$-barrel proteins find that the 
folding temperature is increased and the folding time
decreased~\cite{FriedelM03,CheungMS05}. 
However, molecular crowding has been shown in secretory cells to impair
protein folding and lead to aggregate formation in the
ER~\cite{Ionescu2011}. 
It has been estimated that increasing
the total intracellular protein concentration by 10\% can potentially
increase the rate of protein misfolding reactions following a
nucleation-polymerization mechanism by a factor or 10~\cite{MintonAP2000}.  
Consistent with these observations and estimates, another MD folding study of
a coarse-grained model of crambin found that the presence of multiple
protein copies with a weak 
inter-protein attractive potential (a more realistic scenario)
hindered correct monomeric folding and predisposed the system to
aggregation and misfolding~\cite{WojciechowskiM2008}. 

The above considerations motivate the creation of computational
models, with which we can account for the cellular environment around
a protein in an accurate but less computationally expensive way. 
We begin this paper by reviewing some common methods for
  calculating the free energy to transfer a molecule from one solvent
  environment to another. Two of the most common of these are
  phenomenological continuum approaches and liquid state theory approaches.

The observed linear dependence of the log solubility on the number of CH$_2$ groups and
hence chain length, particularly
for long chain saturated fatty acids (decanoic acid and longer), and 
long-chain aliphatic alcohols (1-butanol and longer), can be taken to
indicate a free energy change upon transfer to solvent that scales
linearly with either volume or surface area. 
Historically, surface area has been taken, under the assumption that
interactions with the solvent take place at the surface of the
molecule in question \cite{LeeB71,EisenbergD86}. Then the free energy
difference between an amino acid in water and in a 
solvent with some osmolyte concentration is, for a given configuration, given in terms of the accessible
surface area (ASA) of that configuration by the phenomenological
expression: $\Delta F = \gamma \cdot \mbox{ASA} + c$, where $\gamma$
is obtained from, {\it eg}, a tri-peptide 
experiment~\cite{TanC2007}. 

The coefficient $\gamma$ depends on the atomic species being
transferred. A more refined approach is thus necessary for a protein, wherein 
the accessible area of the various types of amino acids along with the
backbone are treated differently, so that
\begin{equation}
\label{eq:sasa}
\Delta F = \sum_i \gamma_i \mbox{ASA}_i + c
\end{equation}
The $\gamma_i$ values are taken to be distinct for polar and
non-polar residues, and may even depend on the specific amino acid
identity~\cite{OBrienEP08,AutonM04}. 

Recent simulation studies have found significant
volume contributions to transfer free energies
however~\cite{LinhanantaA11}. In these studies, model solvents with no
enthalpic interaction (hard sphere solvents) still showed significant
transfer free energies, due solely to excluded volume. 
Volume corrections to the surface area model, computed by scaled
particle theory or RISM approaches,  have been
investigated by several
  authors~\cite{SodaK1993solvent,SaundersAJ00,Schellman03,ImaiT2007}.
As well, Baker and colleagues have found that the inclusion of
volume terms (computed by scaled particle particle theory) and 
dispersion integral terms (computed by Weeks-Chandler-Andersen theory)
were essential for an accurate implicit solvent description of atomic-
scale nonpolar forces~\cite{WagonerJA06}.

Obviously, the phenomenological approach can only approximately
capture the effects of the environment, which will include both
interaction energies between the osmolytes and the protein, and terms
arising from the change in entropy of the osmolyte bath. These
techniques, though, are
popular~\cite{QiuD97,RouxB99,BashfordD2000generalized,ZhouR02,GallicchioE04,TanC2007,LabuteP08,ChenJ2008Recent},
computationally cheap to implement, and generalizable to include
  continuum electrostatic and van der Waals terms to accurately parameterize a
  given solvent- typically water~\cite{FeigM04,ChenJ2008Recent}.

Approaches based on liquid state theory generally seek to calculate
the correlation function between sites within the protein and some
model for a continuous medium surrounding it. One approach to doing
this is the reference site interaction model (RISM)\cite{ChandlerD72},
which defines sites in the protein and the surrounding molecules. Once
these sites are defined, the correlation function between them can be
determined using the Ornstein-Zernike
equation~\cite{BarkerJA76}:
\begin{equation}
h( r )  = c( r ) + \int dr^\prime \rho\: c(r^\prime)h(r^\prime-r) 
\label{eq:oz}
\end{equation} 
where $h(r )$ is the total correlation function, $c( r)$ the direct
correlation function, $\rho$ the solvent density. To solve equation
(\ref{eq:oz}), a closure relation is needed, such as the hyper-netted
chain (HNC)
closure~\cite{BarkerJA76,HirataF1981extended,BeglovD1997integral,KovalenkoA99}:
\begin{equation*}
h( r ) = \mbox{e}^{-\beta u( r ) + h( r ) - c ( r )} - 1
\end{equation*}
or the Kovalenko-Hirata closure~\cite{KovalenkoA98}. Here $u( r )$ is the direct interaction potential between particles.
Once correlation functions have been calculated, transfer free energies can be determined through standard methods.

Another liquid state theory approach uses the 
density functional theory (DFT) developed originally for electronic
structure calculations and applies it to condensed classical
systems~\cite{MerminND65thermal,EbnerC76,EvansR79,ChandlerD86a,EvansR92,BibenT98}.
It is noteworthy that
Peter Wolynes has
made significant contributions to the application of density
functional methods in condensed matter systems, primarily
through his fundamental studies of glass 
physics and the glass transition~\cite{SinghY85,Kirkpatrick87a,HallRW87,XiaXY00,XiaXY01,Stevenson2006shapes,LubchenkoV07},
but also in protein
folding\cite{ShoemakerBA97,Shoemaker99b,PortmanJJ01}. 
Density functional theory has also been applied to a variety of non-homogenous
systems such as associating liquids and polymer nanocomposites
\cite{EmborskyCP2011,WuJ2006}. 
Takada and colleagues have used DFT to address crowding effects on the
aggregation of proteins,
wherein protein concentration is treated as a density field with the whole
protein simplified to a sphere~\cite{KinjoAR2002a,KinjoAR2002b}. 

Wolynes's previous applications of DFT to address problems in
disordered condensed matter systems have, along with the other studies
mentioned above, inspired us to continue this tradition in chemical
physics and his legacy in that context, and consider the application
of DFT to protein stability. Here we treat the protein as the source 
of an external potential, which allows a much more realistic protein
model. This approach has certain advantages,
which we will return to later. 

Liquid state theory approaches
have been shown to give solvation densities consistent with values
  from explicit solvent
  calculations\cite{StumpeMC2010calculation}; 
and can be refined to arbitrary accuracy by including additional
  interaction site, three body correlations, and quantum
  corrections.~\cite{BarkerJA1971liquid,IhmG91,HarrisJG92,MasEM03,WangL06}
Liquid state theory 
can be used to determine correlation
functions for the constituent atoms within osmolytes as well as
osmolytes as whole,  so
that effects such as orientation of polar solvents can be
captured. 
Liquid state theories are generally much more accurate than
phenomenological approaches such as
Equation~(\ref{eq:sasa})~\cite{TanC2007},
particularly when discrete molecular aspects of solvation are
  important. The principle disadvantage is the large computational cost
of solving the equations for each configuration.
As well, implicit solvation models using continuum electrostatics 
with optimized parameters (GB/SA) are now capable
  of obtaining solvation energies typically within $\sim$1 kCal/mol of
  experimental values for
  small neutral solutes, while charged solutes tend to show larger
  errors.~\cite{ZhuJ2002parametrization}
Nevertheless, GB/SA continuum methods have shown increased utility
  and widespread use for molecular dynamics
  simulations~\cite{FeigM04,ChenJ2008Recent}.

A large body of literature is concerned with calculating the
electrostatic response of a continuous media to the insertion of a
molecule\cite{CramerCJ99implicit}. This is vitally important in the
context of water solvation. 
In this context, DFT has been applied to the problem of solvation by
Borgis and colleagues~\cite{RamirezR05,BorgisD12}. 
We became aware of their work only in the late in the stages of preparing this
manuscript; our approach is similar at least in spirit to
theirs, however we take a more conceptual approach to address much
larger protein systems and the effects of osmolyte solutions, and how
the DFT framework subsumes many of the notions contained in simplified heuristic
models. 

\begin{figure}
\includegraphics[width=8.6cm]{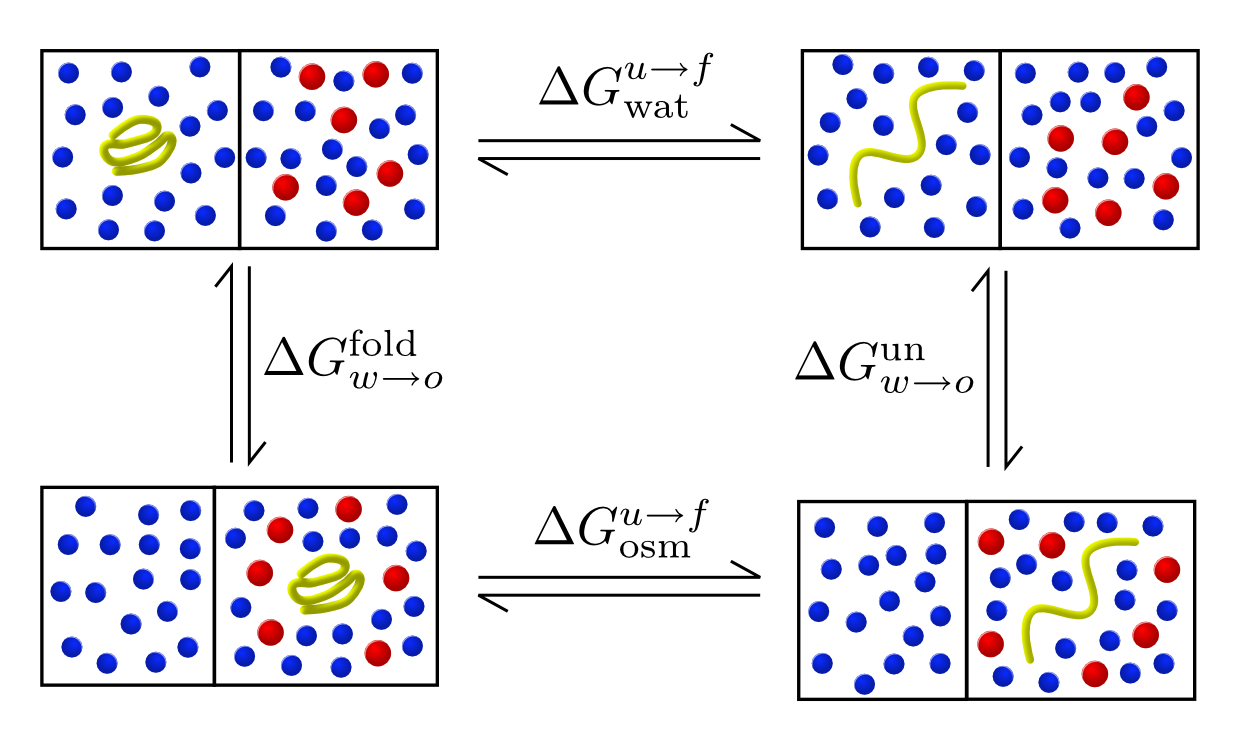}
\caption{A diagram of the Tanford transfer model, for a transfer process going from a pure water environment to one of water and osmolytes. Knowledge of the
  free energy of unfolding $\Delta G^{u\rightarrow f}_{\mathrm{wat}}$
  in the absence of osmolytes can be combined with the transfer free
  energies of the folded ($\Delta G^{\mathrm{fold}}_{w\rightarrow o}$)
  and unfolded  ($\Delta G^{\mathrm{un}}_{v\rightarrow s}$) states to
  obtain the free energy of unfolding in the presence of osmolytes
  $\Delta G^{u\rightarrow f}_{\mathrm{osm}}$.}
\label{fig:tranfer_diag}
\end{figure}

The problem we consider in this paper is that of calculating the free
energy change upon moving a solute such as a protein from a pure
water environment and inserting it into a water and osmolyte
environment. \ref{fig:tranfer_diag} illustrates the problem we are
considering in the context of the Tanford transfer model for
  protein folding~\cite{TanfordC64}; we will return to this diagram several times throughout the paper.

The organization of this paper is as follows. We begin in Section~\ref{sec:gen_con} by 
 investigating the expected behavior of the
 surface and volume 
contributions to the transfer free energy in a 
heuristic model.
In section \ref{sec:dft} we derive the
principle equations for the DFT model of the
transfer free energy. In section
\ref{sec:simp_mod} - section~\ref{sec:imp_solv}, we consider several
examples of how the DFT model can be applied, making connections
  with the model developed in
  Section~\ref{sec:gen_con}. We
  finally conclude and give our outlook on future
directions for this approach.

\section{Volume and Area Terms in the Transfer Free Energy}
\label{sec:gen_con}

\subsection{Volume Considerations}
\label{sec:vol_con}

\begin{figure}
\includegraphics[width=8.6cm]{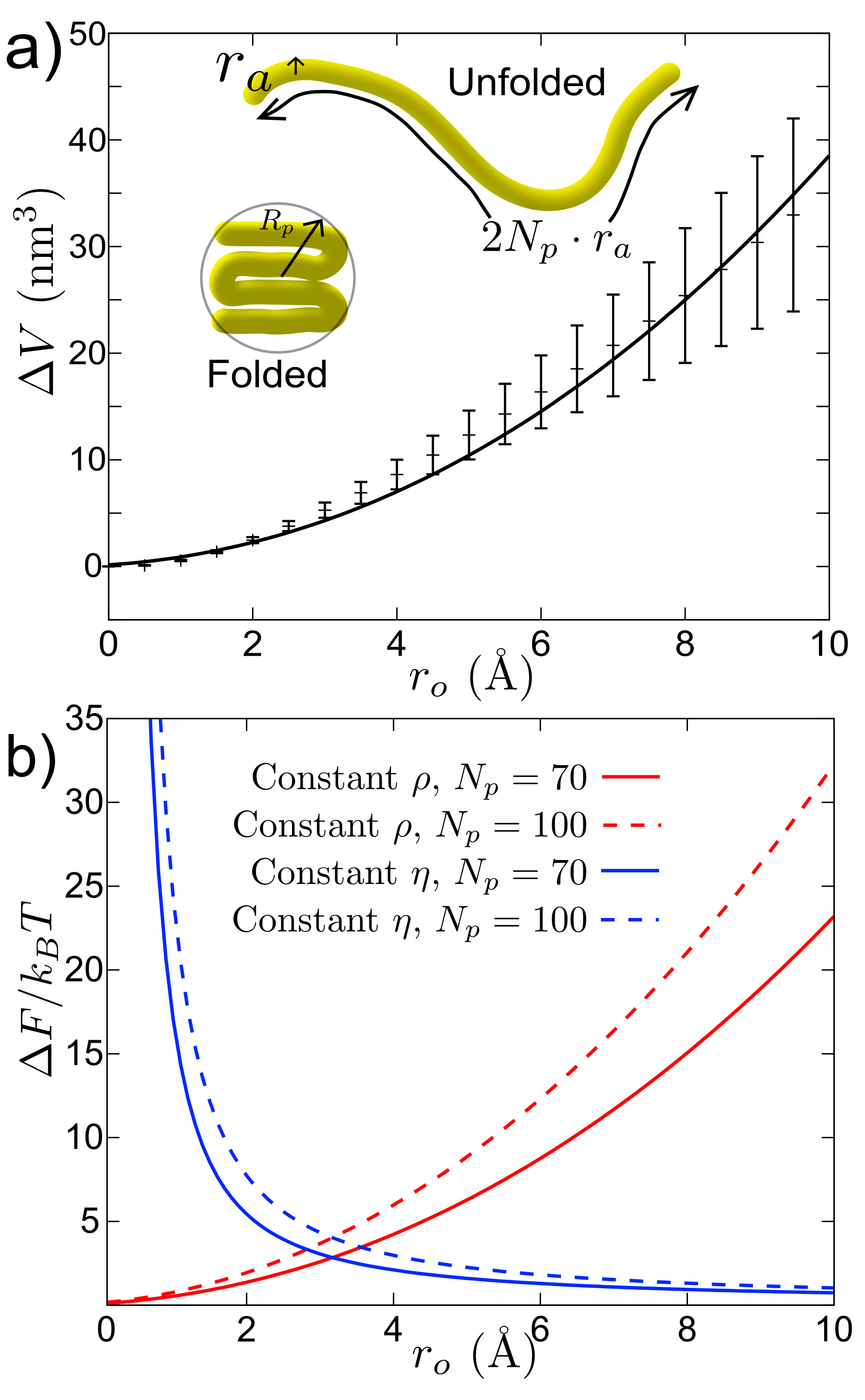}
\caption{
a) Plot of minus the change in volume upon collapse $\Delta V(r_o) = V_u - V_f $,
  as  a function of osmolyte radius $r_o$, for a polymer chain of
  length $N_p=70$ residues  and with $r_a=6$ \AA. The magnitude of the change in volume 
monotonically increases as $r_o$  increases.
Also plotted are the average $\Delta V = \langle V_u\rangle - \langle V_f \rangle$ values of simulation
trajectories of Cold-Shock Protein ($N=70$, PDB 2L15) against probe
radius. (Inset) Schematic of collapsed/folded and unfolded polymer. Folded polymer has radius $R_p$; unfolded polymer has tube radius $r_a$ and length $N_pr_a$. b) Minus the change in free energy upon collapse as a function of osmolyte
  radius $r_o$, for both constant packing fraction $\eta$ and
  constant concentration $\rho$. The
  value of $\rho$ was set to $1M$, and the value of $\eta$ was set so
  that the free energy change would be equal to that at constant $\rho$
at a typical osmolyte
  radius of $3.1$ \AA.  This gave a packing fraction $\eta \approx 0.075$.
}
\label{fig:fvsro}
\end{figure}

To appreciate the terms that we expect in an expression for the transfer free energy, we initially
consider both volume and surface area effects in a more
qualitative way. We consider the difference in volume occupied by the
folded and unfolded states, or more precisely the expanded and
collapsed states of a polymer, to obtain the corresponding free energy
difference in the presence of a bath of ``hard-sphere''
osmolytes. There are thus no surface interactions to consider, and we
seek to estimate the magnitude of the volume effect; we also ignore
for the time being the change in internal free energy as the polymer
collapses. The free energy change  upon collapse of a
protein or polymer then
arises from the change in entropy of the osmolytes, due to the change
in available phase space.  
For hard-sphere osmolytes, the volume occupied by the expanded polymer will
be larger than that of the collapsed polymer. The same considerations
apply to a collapsed {\it vs.} expanded protein; unfolded states
of proteins are generally found to be expanded relative to the
folded state~\cite{KohnJE04}. 
In what follows, let $r_a$ be the mean amino
acid radius, $r_o$ the osmolyte radius, and $N_p$ the number of
amino acids in the polymer or protein. 
Treating the
unfolded protein crudely as a meandering cylindrical tube (see \ref{fig:fvsro}a inset), 
the volume is approximately $\pi
(r_a+r_o)^2 (2 r_a N_p+2 r_o)$, which is that of a cylinder
of radius $r_a + r_o$ and length $2 N_p r_a+2 r_o$.
The volume of the collapsed
globule, or folded protein, can be modelled as a sphere of radius $R_p
+ r_o$, where $R_p$ is the protein radius as probed by a zero-radius
osmolyte particle, i.e. the collapsed volume is $(4/3) \pi (R_p
+r_o)^3$. When $r_o=0$, the unfolded and
folded volumes must be equal, giving $R_p^3 = (3/2) N_p r_a^3$. The
change in available volume for osmolytes $\Delta V (r_o)$
upon polymer collapse is thus positive, and
is plotted in \ref{fig:fvsro} as a function of osmolyte radius $r_o$,
for a chain of length $N_p=70$.

We can compare the results of the above simple model to 
data taken from simulations of a C$_\alpha$ G\={o} model of cold-shock
  protein (PDB 2L15), with $70$ amino acids, generated with the
  GROMACS  molecular dynamics package. The G\={o}
  potential was generated using a shadow map for the native
  contacts~\cite{NoelJK12} by the SMOG@ctbp
  server~\cite{NoelJK10b}.
The simulated free energy surface has a double-well structure with well-defined folded ($f$) and unfolded ($u$)
  ensemble as observed in C$_\alpha$ G\={o} models for other single domain proteins~\cite{ClementiC00:jmb}. We take conformational snapshots in
  each ensemble and measure the volume using a variable probe
  radius with the program VOIDOO~\cite{KleywegtGJ94}. The average volume change $\Delta V = \langle V_u\rangle - \langle V_f \rangle$ for a given probe radius is
  plotted in \ref{fig:fvsro}a. The
theory and simulation data compare quite well given the simplicity of
the model. 

We now consider the free energy as a function of either uniform density $\rho$ or packing fraction $\eta$ of the osmolytes.
Given a large effective box with volume $V_{box}$ containing a given
protein, the packing fraction of osmolytes $\eta$ (i.e. the
volume density) is given by 
$$
\eta = \frac{ \frac{4}{3}\pi r_o^3 N_o}{ V_{box} - V_{prot}(r_o)}
\approx \frac{\frac{4}{3}\pi r_o^3 N_o}{ V_{box} } =\frac{4}{3}\pi r_o^3 \cdot \rho\: ,
$$
where $\rho$ is the number density.
So, at a fixed packing fraction the number of osmolytes $N_o$ scales as
$r_o^{-3}$. 

To estimate the volume contributions to the free energy change upon collapse, $\Delta F_V(r_o)$, as a function of osmolyte radius
but at either fixed density or packing fraction, we use the ideal gas approximation for
the osmotic pressure $p_{osm}=\rho k_BT$ to obtain 
\begin{equation}
\Delta F_V(r_o) = p_{osm} \Delta V (r_o)= \rho k_BT \Delta V (r_o)  = 
\frac{\eta k_B T \Delta V(r_o)}{ \frac{4}{3}\pi r_o^3 }
\label{eq:f_vol}
\end{equation}
where $\Delta V(r_o)$ is obtained from the model above.

A plot of the magnitude of the free energy change upon collapse as a function of
osmolyte radius, here exclusively due to the increase in
entropy of osmolyte particles, is shown in \ref{fig:fvsro}b. 
Based on these considerations we can estimate the volume-like
contribution for typical osmolyte sizes and concentrations. Taking
TMAO as an example, we expect the osmolyte radius to be about 2 \AA, from
the water oxygen-TMAO nitrogen radial distribution
function\cite{KastKM03}. Given this radius and a 
concentration of 300 g/L, for a protein of length $N_p=70$ we  estimate
a volume contribution to the free energy of $ \approx 4k_BT$. 
 The free energy of unfolding is linear in protein
length, so a larger protein of $N_p=300$ has an estimated $\Delta G
\approx 17 k_BT$.

\subsection{Surface Considerations}
\label{sec:sa_con}

The presence of osmolytes in solution can make the effective solvent more
repulsive to protein resulting in stabilization, or more attractive to
the protein resulting in denaturation. What effect is observed depends
on the energy $\eps$ of osmolyte-protein binding and also the
concentration $c$ (or equivalently the chemical potential $\mu$) of the osmolyte.

The energy $\eps$ of binding of the osmolyte is actually the difference in internal
free energy of binding between osmolyte and water, since for example
water may have some attraction to the polymer, and also an osmolyte may supplant more than one
water molecule in the process of binding. 

Previous treatments of transfer free energy analysis as a condensation
problem onto the surface of the protein have been undertaken primarily in the context of protein
denaturation and the prediction of
$m$-values~\cite{SchellmanJA1994thermodynamics,MyersJK95,AlonsoDOV91}. 
The process of condensation of an osmolyte to a surface is equivalent
to the well-known statistical mechanical problem of Langmuir's
isotherm~\cite{HillTL60}, for which the partition function
$\mathfrak{Z}$    
in the $(T,\mu)$ ensemble for a substrate with $M$ absorbing sites is
given by $\left( 1 + \mbox{e}^{-\beta \left( \eps - \mu \right)}
\right)^M $.  The mean covering ratio $f$ is then given by 
\begin{equation}
f = \frac{kT}{M} \frac{\partial \log \mathfrak{Z} }{\partial \mu} =
\frac{1}{1+ \mbox{e}^{\beta \left( \eps - \mu \right)} } \: ,
\label{eq:f}
\end{equation}
and the mean energy of condensation on the surface is $M f \eps$. Here
we neglect interactions between osmolytes when bound.  
The Helmholtz free energy in this model is given by 
\begin{equation*}
F = -p V + f M\mu = -k_BT\log(\mathfrak{Z}) + f M \mu 
\end{equation*}
with $T,\mu$ partition function $\mathfrak{Z}$ as given above. 

We can relate the Langmuir isotherm to the free energy of a protein
surface by assuming that each osmolyte occupies an area $a_0 \approx
\pi r_o^2$ on the protein surface, so that we can write $M=A/a_0$,
where $A$ is the protein's solvent accessible surface area in a given
conformation. 
The change in free energy $F_A$ upon condensation becomes 
\begin{equation}
F_A = -k_BT \frac{A}{a_0}
\log\left(1+\mbox{e}^{-\beta \left( \eps - \mu \right)} \right)  
+  f \frac{A}{a_o} \mu
\label{eq:f_area}
\end{equation}

If the concentration of unbound osmolyte is dilute, an ideal gas approximation
suffices for the chemical potential: $\mu = kT \log \left(
  \rho/\rho_Q\right)$, where $\rho_Q$ is a reference concentration (typically taken to be 1M).
The quantity $\mathrm{e}^{-\beta \eps}/\rho_Q$ is typically treated as
an equilibrium constant in the
literature~\cite{MyersJK95,AlonsoDOV91}.
We consider both dilute and non-dilute limits below.
The protein's exposed surface area is obtained from the volume given
in section~\ref{sec:vol_con}  
by $A = \partial
V/\partial r_o$, so the collapsed exposed area is $4 \pi \left(R_p + r_o
\right)^2$ and the expanded (random coil) exposed area is $2 \pi
\left( r_o + r_a\right) \left[ \left( 2 N_p + 1 \right) r_a + 3 r_o
\right]$. 

\subsection{Combined surface/volume model for the transfer free energy}
\label{sec:combined} 

We can now write the total free energy of collapse $\Delta F$ arising from
osmolytes by combining the volume and surface area terms in equations
(\ref{eq:f_vol}) and (\ref{eq:f_area}). We can also remove the ideal
  gas assumption by expressing $\Delta F$ in terms of the
  Carnahan-Starling (CS) approximations to the pressure and chemical
  potential:~\cite{CarnahanNF1969equation}
  \begin{eqnarray}
  p &=& \rho \kB T \frac{1+\eta+\eta^2-\eta^3}{(1-\eta)^3} \nonumber \\
  \mu &=& \kB T \log(\rho/\rho_Q) + k_BT
  \frac{8\eta-9\eta^2+3\eta^3}{(1-\eta)^3}  \: .
\label{eq:pmucarnahan}
  \end{eqnarray} 
Then the free energy becomes:
\begin{equation}
\Delta F = p \Delta V + \left( \frac{\kB T}{\pi
  r_o^2}\log\left( 1-f \right) 
+ \frac{f \mu}{\pi r_o^2} \right)
\Delta A
\label{eq:simp_con_total}
\end{equation}
with $f$ given in~(\ref{eq:f})
and $p$ and $\mu$ given in~(\ref{eq:pmucarnahan}),
and where
\begin{align*}
\Delta V(r_o) &=
\frac{4}{3}\pi\left(\left(\frac{3N_p}{2}\right)^{1/3}r_a+r_o\right)^3
- 2\pi(r_a+r_o)^2(N_pr_a+r_o) \\
\Delta A(r_o) &=
4\pi\left(\left(\frac{3N_p}{2}\right)^{1/3}r_a+r_o\right)^2 -
2\pi(r_a+r_o)[(2N_p+1)r_a+3r_o]
\end{align*}
are the volume and surface area change upon folding (or
  collapse). 

\begin{figure}
\includegraphics[width=8.6cm]{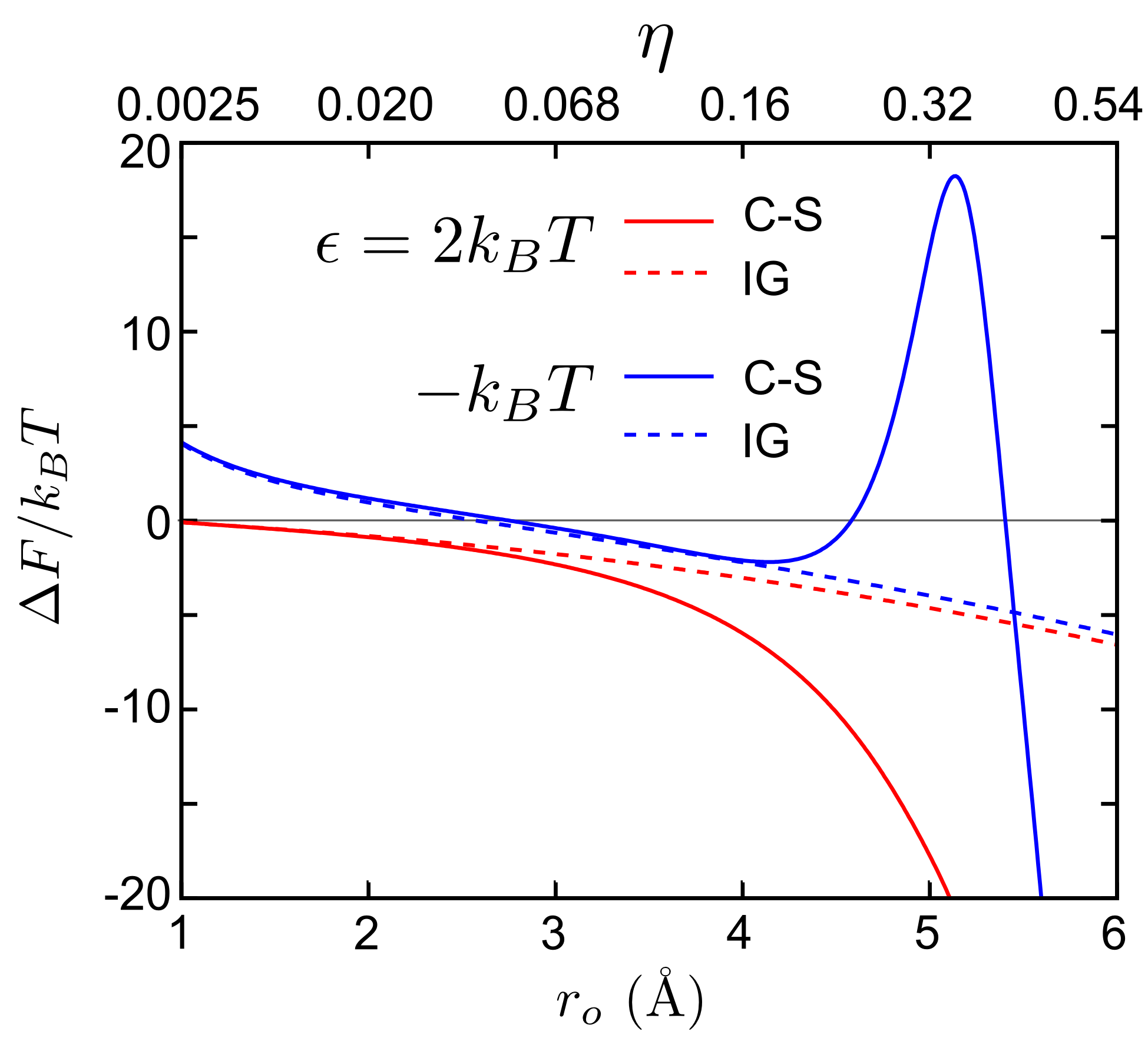}
\caption{
Total free energy change $\Delta F$ upon collapse
in units of $\kB T$, as a function of osmolyte radius $r_o$.
Values of packing fraction $\eta$ corresponding to the values of
  $r_o$ on the x-axis are shown above the plot.
Curves are taken from equation (\ref{eq:simp_con_total}) which combines surface area
and volume terms. 
Here the polymer length $N_p=70$, the osmolyte concentration
$\rho=1$M, and $r_a=6$\AA.  
Red curves show $\Delta F$ upon collapse for a repulsive osmolyte
  with interaction energy $+2\kB T$,
i.e. a crowding particle. Blue curves show $\Delta F$ upon collapse
for an attractive osmolyte  with interaction energy $-\kB T$, i.e.  a weak denaturant.
Plotted are both the model with ideal gas (IG, dashed) and 
Carnahan-Starling (C-S, solid) 
pressure and  chemical potential.  
}
\label{fig:simp_cond_total}
\end{figure}

We plot
equation 
(\ref{eq:simp_con_total}) in \ref{fig:simp_cond_total} as
  a function of osmolyte radius $r_o$, for condensation energies $\epsilon=2k_BT$ and
  $\epsilon=-k_BT$. To assess the limits of the ideal gas model, we
  have also plotted the ideal gas results in \ref{fig:simp_cond_total}. 
For repulsive osmolyte-protein interactions, both surface and volume
terms stabilize the folded or collapsed state
(\ref{fig:simp_cond_total}).
The free energy change upon collapse is 
monotonically decreasing (increasing in magnitude) from zero, and more
strongly favoring collapse as
osmolyte radius is increased. Non-ideal excluded volume effects in the
osmolyte pressure and chemical potential enhance the stabilizing effect. 
For attractive osmolyte-protein interactions, the situation is more
complex. At small values of osmolyte radius $r_o$, the collapsed phase
is destabilized by osmolyte-protein binding, which favors expansion. 
As $r_o$ increases, the volume change upon
collapse both increases, which begins to entropically favor
collapse. The osmotic pressure initially increases modestly,
additionally favoring collapse. However the chemical potential also
increases modestly, driving condensation of osmolyte and favoring
expansion. These two effects nearly cancel each other rendering the
real and ideal gas curves nearly coincident up to $r_o\approx
4$\AA.
The sigmoidal dependence of covering fraction $f$ 
in equation~(\ref{eq:f}) on chemical
potential $\mu$ results in a sudden condensation of osmolyte onto the
protein around $r_o \approx 5$\AA, which induces the system to favor
expansion at these radii. While the number of condensed osmolytes is
bounded, the osmotic pressure is not, and eventually collapse is
favored once again through volume terms. 
The osmolyte radius $r_o$ can
only increase until $\eta \approx 0.6$ (near crystal packing
densities), giving a cutoff of $r_o^{(cut)} 
\approx (3\eta/4\pi\rho)^{1/3}$, or about $6.2$\AA~for 1M
concentration. 

In the limit that the osmolyte is dilute,
$
\rho\mathrm{e}^{-\beta\eps}/\rho_Q \ll 1
$
and we can expand the logarithm in equation (\ref{eq:simp_con_total})
to obtain an area contribution to the free energy of 
$ -\rho \kB T A \mathrm{e}^{-\beta \eps} /a_0\rho_Q $,
so that the free energy change upon  unfolding becomes
\begin{equation}
\Delta F = \rho \kB T \left(  \Delta V 
- A t \mathrm{e}^{-\beta \eps} \right) \: .
\label{eq:approx_A}
\end{equation}
Here we have used the fact that $(a_0\rho_Q)^{-1}$ has units of
length and can be thus be interpreted physically as a thickness $t$ over which the surface interaction acts.

Having looked at these preliminary volume and surface considerations,
we now turn to a classical density functional theory formulation,
which provides a more complete understanding of the transfer free
energy, and as well, reduces to equation~(\ref{eq:approx_A})
in the appropriate limits.

\section{The Density Functional Theory Formulation}
\label{sec:dft}

We now consider a density functional formulation of the problem of
  transfer free energy. In what follows, we will assume that the
intra-protein energy of a given configuration of a protein is in
principle known and the net interaction between any given site on the
protein and either the osmolyte or water is in principle known. We
then wish to calculate $\Delta F$, the free energy of transferring the
protein from water to an osmolyte solution, or, equivalently, of
transferring the osmolytes from an aqueous solution to one containing
the protein (see \ref{fig:tranfer_diag}). In short, we wish to
consider the effect that the presence of osmolytes has on the free
energy of the protein. 

The uniqueness of the Kohn-Sham density functional may be extended to finite temperatures, so that the free energy of the protein-solvent system is uniquely expressed as a functional of the single particle density $\phi(\bm{r})$~\cite{PlischkeM}. We thus seek an expression for the free energy of the osmolytes and water in an arbitrary external potential. For our purposes in obtaining a transfer free energy, we will treat a given protein configuration, with atom positions $\{ \bm{R}_i \}$, as the source of the external potential. We write the free energy in the standard way~\cite{EmborskyCP2011}:
\begin{align}
F(\{ \bm{R}_i \}) = & \int d^3r  \: k_B T (-S_o(\phi_o(\bm{r})) - S_w(\phi_w(\bm{r}))) + \mathcal{V}_o(\bm{r})\phi_o(\bm{r}) + \mathcal{V}_w(\bm{r}) \phi_w(\bm{r})   \nonumber \\
& + \Phi_o[\phi_o] + \Phi_w[\phi_w] + \Phi_{ow}[\phi_o,\phi_w]
\label{eq:dft_full}
\end{align}
Here $\phi_j$ is the density function for the osmolytes $(o)$ or water
$(w)$, and $\mathcal{V}_j$ the external potential on the respective species.
The entropy density
for each species can be written as
\begin{equation}
\label{eq:S}
S_o(\bm{r})+ S_w(\bm{r}) =  
- \phi_o( \bm{r} ) \log \left[  \lambda_o^3\phi_o(\bm{r}) \right] - 
\phi_w( \bm{r} ) \log \left[  \lambda_w^3\phi_w(\bm{r}) \right]
\end{equation}
where $\lambda_o$ and $\lambda_w$ are constants with units of length,
analogous to thermal wavelengths.
The terms $\Phi_o$, $\Phi_w$, and $\Phi_{ow}$ are the
multi-particle correlation contributions to the free energy for the
respective species. For example, the two particle correlation part of
$\Phi_o$ would have the form 
\begin{equation}
\Phi_o^{(2)}[\phi_o] = \int \int d^3r_1 d^3r_2 \: \phi_o(\bm{r}_1) \phi_o(\bm{r}_2) U_{oo}(\bm{r}_1-\bm{r}_2) g(\bm{r}_1,\bm{r}_2|\mathcal{V})
\label{eq:two_corr}
\end{equation}
where $U_{oo}$ is the interaction potential between two osmolytes and $g$ the two-particle correlation function. The full multi-particle function is not known exactly, and so, as in electronic DFT, while equation (\ref{eq:dft_full}) is exact in principle, approximations must be made to use it in practice ~\cite{JonesRO89}.

We now make two key assumptions. The first is that the water and
osmolyte densities are completely correlated, such that all vacua are
occupied by either water or osmolyte. Thus $N_w v_w + N_o v_o = V$,
where $v_i$ is the volume of an individual water or osmolyte molecule,
and $V$ the total volume. Dividing this by $V v_w$ and allowing
  the local
density of a given species to vary gives
\begin{equation}
\phi_w(\bm{r}) + f \phi_o(\bm{r}) = \rho_w
\label{eq:fixedv}
\end{equation}
where $f = v_o/v_w$ and $\rho_w = 1/v_w$ (the factor of $f$ allows for
the osmolyte molecule to be a different size than the water
molecule). Equation~(\ref{eq:fixedv})
is not valid in the interior of the protein, so we split our system up
into two regions: a hard wall region $V_{hw}$ in which
$\phi_w=\phi_o=0$, and the rest of the system, which has a volume $V$
identical to the volume of the osmolyte-water bath prior to the
insertion of the protein, and in which Equation~(\ref{eq:fixedv}) is valid.
We further take $V_{hw}$ to be same as the
change in volume of the aqueous system the protein was removed from in
the transfer process (see  \ref{fig:tranfer_diag}), so that the total system of water, protein, and osmolyte-water solution does not change volume during the transfer process.

With the approximation of equation (\ref{eq:fixedv}) we can write
\begin{align}
\mathcal{V}_o(\bm{r})\phi_o(\bm{r}) + \mathcal{V}_w(\bm{r}) \phi_w(\bm{r}) = & \Delta \mathcal{V}(\bm{r}) \phi_o(\bm{r})+ \mathcal{V}_w(\bm{r}) \rho_w \label{eq:delv_def} \\
 \Phi_o[\phi_o] + \Phi_w[\phi_w] + \Phi_{ow}[\phi_o,\phi_w] =& \Phi_{t}[\phi_o]
\end{align}
where $\Delta \mathcal{V}(\bm{r}) = \mathcal{V}_o(\bm{r}) - f \mathcal{V}_w(\bm{r}) $.

The second approximation in our treatment is that the osmolyte number density is much
less than that of water. 
Using this approximation along with the one given in Equation~(\ref{eq:fixedv}),
the entropy in Equation~(\ref{eq:S}) becomes
\begin{align}
-S_o(\bm{r}) - S_w(\bm{r}) =& \phi_o(\bm{r}) \log \left[ \lambda_o^3\phi_o(\bm{r}) \right] + (\rho_w - f\phi_o(\bm{r}))\log\left[ \lambda_w^3(\rho_w -f\phi_o(\bm{r}))\right]  \\
 \approx &  \phi_o(\bm{r}) \log \left[\lambda_o^3\phi_o(\bm{r})\right]  - f \phi_o(\bm{r}) + \rho_w\log\left[\lambda_w^3\rho_w\right] - f \phi_o(\bm{r}) \log\left[\lambda_w^3\rho_w\right] \nonumber 
\end{align}
In this way we express each part of equation (\ref{eq:dft_full}) in terms of osmolyte density and constant terms.
The free energy functional may then be written as
\begin{align}
F =& \int d^3r\: k_BT (\phi_o(\bm{r}) \log\left[\lambda_o\phi_o(\bm{r})\right] - (\gamma+1)
\phi_o(\bm{r})) + \Delta \mathcal{V}(\bm{r}) \phi_o(\bm{r}) \nonumber \\
 &+ V\rho_w  \log\lambda_w^3 \rho_w + \mathcal{U}\rho_w + \Phi_t[\phi_o]
\end{align}
where $\mathcal{U} \equiv \int d^3r \mathcal{V}(\bm{r})$, and $\gamma+1 \equiv f(1+\log(\lambda_w^3\rho_w))$. Since $V$ is the volume of the system, the term $V\rho_w$ is equal to $V/v_w=N_w^\prime$, the total number of water molecules in a system of pure water of volume $V$.

Thus, dropping the subscripts, letting $\mathcal{V} \equiv \Delta \mathcal{V}$, and ignoring any position independent terms, we can write the free energy as
\begin{align}
F =& \int d^3r \: k_B T\left(\phi(\bm{r}) \log \lambda^3 \phi(\bm{r}) -\phi(\bm{r})\right) + k_BT \gamma \phi(\bm{r}) + \mathcal{V}(\bm{r}) \phi(\bm{r}) \nonumber \\ 
 & + \Phi[\phi]
\end{align}
where $\Phi[\phi]$ is the functional containing the multi-particle
correlation part of the free energy, and $\lambda \equiv \lambda_o$ is a constant with units of length analogous to the thermal wavelength, whose value will be shown to be unimportant. For now we will formally manipulate $\Phi$ without making assumptions about its form. We can find the density that minimizes the free energy by use of the Euler-Lagrange equations, with the constraint that the osmolyte density when integrated over the total volume is the total number of osmolytes:
\begin{equation}
\int_V d^3r \: \phi(\bm{r}) = N_o  \: .
\label{eq:dft_constraint}
\end{equation}
We thus write 
\begin{align}
\frac{\delta}{\delta \phi} \left[ F - \mu_o \left(\int_V d^3r \: \phi(\bm{r}) - N_o \right) \right] &= 0 \nonumber \\
\mbox{or} \;\;\;\;\; k_B T\log \lambda^3 \phi(\bm{r}) + \mathcal{V}(\bm{r}) - k_B T \gamma + \frac{\delta \Phi}{\delta \phi} - \mu_o &= 0
\label{eq:solve_el}
\end{align}
where $\mu_o$ is the Lagrange multiplier corresponding to the constraint in equation (\ref{eq:dft_constraint}). Physically, we can interpret equation (\ref{eq:solve_el}) as a statement that $\frac{\delta F}{\delta \phi}$ is equal to the chemical potential $\mu_o$, and thus must be a constant value at all points in space. Solving this for the density field gives
\begin{equation}
\phi(\bm{r}) = \mathrm{e}^\gamma \lambda^{-3}\mathrm{e}^{ -\beta (\mathcal{V}(\bm{r}) + \Phi^\prime - \mu_o)}
\label{eq:phi_start}
\end{equation}
where $\Phi^\prime \equiv \frac{\delta \Phi}{\delta \phi}$. 

To obtain $\mu_o$ from equation (\ref{eq:phi_start}), we use the constraint on the total number of particles in equation (\ref{eq:dft_constraint})
which yields
\begin{equation}
\mathrm{e}^{\beta \mu_o} = \frac{\mathrm{e}^\gamma \lambda^3 N_o}
{\int_{V} d^3r \: \mathrm{e}^{-\beta(\mathcal{V}(\bm{r}) +
    \Phi^\prime)}} \: .
\end{equation}

From here we can obtain the transfer free energy, which is given by the free energy of the osmolyte bath in the presence of the external protein potential, $\mathcal{V}(\bm{r})$, minus the free energy of the osmolyte bath without the protein potential $(\mathcal{V}(\bm{r}) = 0$). We thus have
\begin{align}
\Delta F =& \Delta \mu_o N_o \nonumber \\
 =& - k_B T N_o \log \left(\frac{\mathrm{e}^\gamma \lambda^{-3}}{N_o} \int_{V} d^3r \: \mathrm{e}^{-\beta (\mathcal{V}(\bm{r}) + \Phi^\prime_f(\bm{r}))}\right) \nonumber \\
  & + k_B T N_o \log \left(  \frac{\mathrm{e}^\gamma \lambda^{-3}}{N_o} \int_{V} d^3r \: \mathrm{e}^{-\beta \Phi^\prime_i} \right)
 \label{eq:mu_diff}
\end{align}
where the volume $V$ integrated over is the volume outside of
hard-wall volume of the protein, and is the same in the initial and final systems. The difference $\Delta F$ is independent of $\lambda$ and $\gamma$. 

The bath in the initial state is homogeneous and isotropic, so
$\Phi^\prime_i$ in equation (\ref{eq:mu_diff}) is independent of
position. Thus it may be factored out of the integral, 
\begin{equation*}
\int_{V} d^3r \: \mathrm{e}^{-\beta \Phi^\prime_i} = V \mathrm{e}^{-\beta \Phi^\prime_i}
\end{equation*}
so that
\begin{equation}
\Delta F = -k_B T N_o \log \left( \frac{1}{V} \int_{V} d^3r \: \mathrm{e}^{-\beta(\mathcal{V}(\bm{r}) + \Delta \Phi^\prime)}   \right)
\label{eq:premain}
\end{equation}
where $\Delta \Phi^\prime = \Phi^\prime_f(\bm{r}) - \Phi^\prime_i$.
The expression in equation (\ref{eq:premain}) consists of the
logarithm of the integral of a Boltzmann weight for the effective
potential $\mathcal{V}(\bm{r}) + \Delta \Phi^\prime(\bm{r})$. Here
$\mathcal{V}(\bm{r})$ and $\Delta \Phi^\prime (\bm{r})$ enter on equal
footing. Recall that $\mathcal{V}$ is the protein-osmolyte potential,
treating the protein as an external source. $\Phi^\prime$ is the
functional derivative of the multi-particle part of the free
energy. If we use the two-particle osmolyte contribution from equation (\ref{eq:two_corr}), we obtain
\begin{equation}
\Delta \Phi_o^{(2)\prime} =
 \left. \frac{\delta \Phi_o^{(2)}}{\delta \phi_o(\bm{r})}
 \right|_\mathcal{V}  -  \left. \frac{\delta \Phi_o^{(2)}}{\delta
     \phi_o(\bm{r}) }\right|_{\mathcal{V}=0} = \int d^3r^\prime \:
 \left[ \phi_{of}(\bm{r}^\prime) g(\bm{r},\bm{r}^\prime | \mathcal{V}) -
   \phi_{oi}(\bm{r}^\prime) g(\bm{r},\bm{r}^\prime | \mathcal{V}=0)
 \right] U_{oo}(\bm{r},\bm{r}^\prime) \: ,
\end{equation}
which gives the difference of two terms in the presence and absence of
the external protein potential, where each term corresponds to the
equilibrium-averaged interaction energy between 
osmolytes, up to pair correlations.
Thus the term $\Delta \Phi^\prime$ in Equation~(\ref{eq:premain})
can be interpreted as the change in
energy due to redistribution of the environment in response to the
change in external potential. 

We can recast equation~(\ref{eq:premain}) into a form that will be somewhat more useful later:
\begin{equation}
\Delta F = -k_B T N_o \log \left( 1 + \frac{1}{V}\int_V d^3r \: [\mathrm{e}^{-\beta(\mathcal{V}(\bm{r}) + \Delta \Phi^\prime)} - 1] \right)
\label{eq:main}
\end{equation}
which has the advantage that when $\mathcal{V}$ and $\Delta
\Phi^\prime$ are both zero, the integrand is also zero, and thus the integral can be taken over all space.

In equation (\ref{eq:main}) we can take the limit $V\rightarrow \infty$, with $N_o/V = \rho$ fixed.
Then, assuming that the region over which the integrand in
equation~(\ref{eq:main}) is non-zero is finite, we can
expand the logarithm to first order to obtain
\begin{equation}
\Delta F = -k_B T N_o \frac{1}{V} \int d^3r \: \left( \mathrm{e}^{-\beta(\mathcal{V}(\bm{r})+\Delta \Phi^\prime)} - 1\right)
\label{eq:large_v}
\end{equation}
which has the form 
\begin{equation*}
\Del F = p_{id} \Delta V_{\mathrm{eff}}
\end{equation*}
where $p_{id} = N_o \kB T/V$ is the ideal gas osmotic pressure, and 
$V_{\mathrm{eff}}  = \int d^3r \: \left[ 1-\mathrm{e}^{-\beta(\mathcal{V}(\bm{r})+\Delta \Phi^\prime)}\right]$
is an effective change in volume. In the dilute limit, the osmotic
pressure $p=p_{id}$; then $V_{\mathrm{eff}}$
may be interpreted as the
change in volume available to the osmolytes.

We now need to address $\Delta \Phi^\prime$ to progress further. The
obvious first approximation is to set $\Delta \Phi^\prime =0$; we will
see below that this approximation can in fact go quite a long way,
depending on the solvent. 
This is consistent with the observations in \ref{fig:simp_cond_total}
where the ideal gas approximation, which neglects osmolyte-osmolyte
  correlations, holds for typical molecular radii at 1M concentration.
It is worth noting that this is not ignoring the osmolyte-osmolyte, osmolyte-water, and water-water correlations completely; it is merely assuming that they are the same in the initial and final baths. Making this approximation, we have
\begin{equation}
\Delta F = -k_B T N_o \log \left( 1 + \frac{1}{V}\int_V d^3r \: [\mathrm{e}^{-\beta\mathcal{V}(\bm{r})} - 1] \right)
\label{eq:first}
\end{equation}

Equation~(\ref{eq:first}) represents an approximation to the transfer
free energy that, while severe, nonetheless takes into account both
the change in energy and change in entropy of the osmolyte bath.

\subsection{Validation tests in model solvents}
\label{sec:valid}

\begin{figure}
\includegraphics[width=8.6cm]{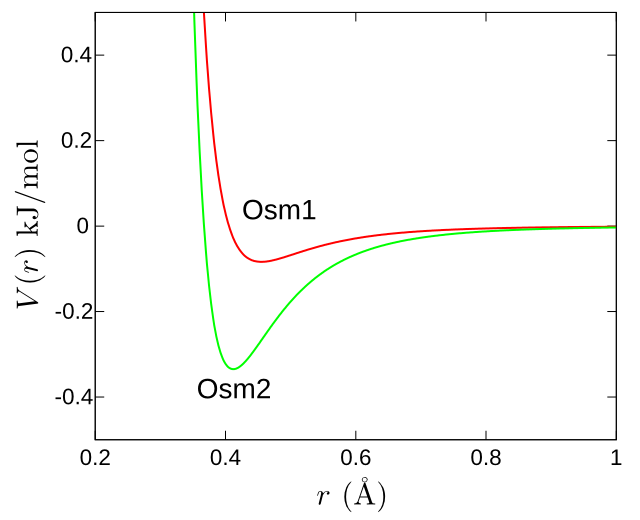}
\caption{Comparison of osmolyte potential functions for the test cases
  parameterized in \ref{tab:osm_par}.  Osm2 is significantly more attractive than Osm1, which is reflected in the transfer free energies in \ref{tab:compare}}
\label{fig:osm_pot_comp}
\end{figure}

As a test of the density functional theory, we have
used equation (\ref{eq:first}) to calculate the transfer free energy
of several small molecules into model osmolytes. To simplify the
simulations, we looked at transfer from vacuum to a van der Waals gas
of osmolytes, which were taken to be single atoms interacting through
a VDW potential. The density of the osmolytes was set to 1M. The
molecules we transferred were the side chains of alanine and valine,
with C-$\beta$ capped with a hydrogen to replace the backbone ({\it
  ie}, the molecules were methane and propane). The coordinates were
taken from an existing protein structure file, and the angle and bond
parameters were generated with the GROMACS utility pdb2gmx. The
charges were set to zero for all atoms, and the interaction was purely
van der Waals. We list the VDW parameters
in \ref{tab:osm_par}. \ref{fig:osm_pot_comp} shows the interaction
potential for the two different osmolytes we used.  The transfer
energies were calculated both with equation (\ref{eq:first}) and by
simulating the transfer in GROMACS and using Thermodynamic
Integration (TI)~\cite{SlusherJT99,WescottJT02,ShirtsMR03}. The results are
summarized in \ref{tab:compare}, and show excellent agreement between
TI and DFT. This is notable since the result was obtained neglecting
the inter-particle correlations, and at 1M the pressure of the
osmolytes was $\approx 1.5$ that of the ideal gas pressure, which indicates that the osmolyte-osmolyte interactions were significant.

\begin{table}
\centering
\begin{tabular}{c|c|c}
Atom & $\sigma$ (\AA) & $\epsilon$ (kJ/mol) \\
\hline 
Ala C-$\beta$       & 0.36705 & 0.33472 \\
\hline 
Ala H                      & 0.23520 & 0.092048 \\
\hline
Val C-$\beta$       & 0.40536 & 0.08368 \\
\hline
Val C-$\gamma$ & 0.36705 & 0.33472 \\
\hline 
Val H                      & 0.23520 & 0.092048 \\
\hline
Osm1                    & 0.40536 & 0.08368 \\
\hline
Osm2                     & 0.36705 & 0.33472 \\
\end{tabular}
\caption{van der Waals parameters for the atoms used in the simulation
  test of the DFT, as taken from the CHARMM parameter set. Osm2 is a relatively attractive spherical
  osmolyte, while the potential of Osm1 is dominated by steric
    repulsion. The interaction is parameterized as $V( r ) =
  4\epsilon \left[ \left(\sigma/r \right)^{12} -
    \left(\sigma/r\right)^{6} \right]$.
    }
\label{tab:osm_par}
\end{table}

\begin{table}
\centering
\begin{tabular}{c|c|c}
Molecule/Osmolyte & DFT $\Delta G$ (kJ/mol) & TI  $\Delta G$ (kJ/mol) \\
\hline
Ala/Osm1 & $0.188 \pm 0.002$ & $0.187 \pm 0.002 $\\
\hline
Val/Osm1 & $ 0.255 \pm 0.004 $ & $ 0.261 \pm 0.004  $ \\
\hline
Ala/Osm2 &  $0.055 \pm 0.002$ & $0.059 \pm 0.003$ \\
\hline
Val/Osm2 & $-0.018 \pm 0.004$ & $-0.011 \pm 0.004$ \\
\hline
\end{tabular}
\caption{Comparison of test cases between density functional theory (DFT) and thermodynamic
  integration (TI)}
\label{tab:compare}
\end{table}

\subsection{Connecting DFT to previous
    surface/volume models}
\label{sec:simp_mod}

We now take a simplified model of a protein potential to compare with
the results obtained previously in Section~\ref{sec:gen_con}
for the solvent contribution to the change in free
energy upon protein collapse. In this model we will consider the protein to
have an excluded volume of $V_{prot}$; that is, within that volume the
potential is infinite. From the discussion in sections~\ref{sec:vol_con}-\ref{sec:dft}
 concerning excluded volume, we saw that the changes in
volume treated there are volumes from which osmolytes are excluded. 
We also
consider the protein to have a surface region of thickness $t$ that
exerts a potential on the osmolytes of depth $\epsilon$; this region is
sufficiently thin that we can approximate its volume as
$V_{surface}\approx tA$. If we use this model in the expression for
the free energy in the limit of large system size (equation
(\ref{eq:first}) ) then we obtain a free energy upon transfer of
\begin{equation}
\Delta F = \rho \kB T \left(V_{prot} + (1-\mathrm{e}^{-\beta
    \epsilon})tA \right) \: .
\label{eq:simplF}
\end{equation}

The DFT transfer free energy with this simplified model provides a
natural split between the volume contribution $p_{id} V_{prot}$ and the
surface area contribution $p_{id} (1-\mathrm{e}^{-\beta
  \epsilon})tA$. Thus the DFT result, in the appropriate model,
naturally generates the free energy contributions derived in section
\ref{sec:gen_con} from more 
bespoke considerations. Specifically, if we 
take the total volume of the protein upon insertion to be $V=V_{prot}+tA$,  
then equation (\ref{eq:simplF}) is
identical to equation (\ref{eq:approx_A}). 
 The simplified DFT model here reduces to our earlier
considerations and helps give a physical interpretation of the
  quantity $\rho_Q$ as
it pertains to the protein surface.

We can also see that, in order to obtain an SASA approximation in which $\Delta
F$ is independent of temperature, one would have to assume 
that  the osmolyte-protein binding energy $\epsilon \ll k_B T$, and
that volume terms were either negligible compared to surface terms, or
they were proportional to them. 
We find below that $\eps \approx \kB T$ in order to obtain
empirically-derived transfer
free energies to TMAO,  which does not satisfy the above inequality.  
As well, we can use the tube model from Section \ref{sec:gen_con} for
protein volume and surface area to
estimate the relative contributions of volume and area: for an
osmolyte of radius $r_o=2.5$ \AA \: and a protein with $N_p=70$,
$V/tA=0.62$ in the unfolded state, and $V/tA=0.77$ in the folded state. The
volume here is by no means negligible. 

We thus expect on general grounds that the transfer free energy will be
dependent on temperature. One way of looking at the simplified limit
for the transfer free energy in equation~(\ref{eq:simplF})
is as a derivation of a new phenomenological form for the transfer
energy, containing both temperature and volume dependence:
\begin{equation}
\Delta F = \gamma_1 \kB T (V_{solute} ) + 
\gamma_2 \kB T (\mbox{ASA}) \mathrm{e}^{-\beta \epsilon} \: ,
\end{equation}
where one can now fit the parameters $\gamma_1$, $\gamma_2$, and $\epsilon$, to empirical
data. 

\section{Empirically Deriving DFT Transfer Free Energy Parameters}
\label{sec:emp_params}

The potential $\mathcal{V}(\bm{r})$ in equation~(\ref{eq:first}) is an
effective potential given by $\mathcal{V}_o(\bm{r}) - f \mathcal{V}_w(\bm{r})$. Obtaining
$f$ and $\mathcal{V}_w$ may be nontrivial to obtain {\it ab initio},
so we examine some model systems, and compare with empirical methods. 
To begin with, we will assume that the
potential takes the form of a sum of terms from each particle in the
protein, where a particle may be an atom in an all-atom model, or a
bead modeling an amino acid in a coarse-grained approach: 
\begin{equation*}
\mathcal{V}(\bm{r}) = \sum_{i=1}^{N_p}
v^{\mathrm{eff}}_i(\bm{r}-\bm{R}_i) \: .
\end{equation*}
Here $N_p$ is the number of particles in the protein, and $\bm{R}_i$
the position of the $i$th particle.

We consider a model consisting of backbone $C_\alpha$ atoms and
coarse-grained side-chain beads, which then form the particles for our
potential. We make the assumption that the protein-osmolyte potentials 
have a 6-12 form: 
\begin{equation*}
v_i( r ) = 4 \eps_i \left[ \left(\frac{\sigma_i}{r}\right)^{12} -
    \left(\frac{\sigma_i}{r}\right)^6 \right] \: ,
\end{equation*}
and we wish to determine the potential parameters
$\sigma_i$, $\eps_i$ for each amino acid that reproduce the 
transfer energies found experimentally when DFT is applied using the
above potential. As a
starting point, we examine those used by Auton and
Bolen~\cite{AutonM05}.

Two constraint equations are required for each amino acid. For the
  first equation, we note that
the beads representing the various amino acid side chains have residue
radii $r_{oi}$ that may be obtained from measured partial molar
volumes~\cite{ZamyatninAA84}. We can then
apply a constraint to the above 6-12 parameters $\sigma_i$, $\eps_i$ by requiring that at a
distance $r_{oi}$ from the residue centre, 
\begin{equation}
\label{eq:vroi}
v_i(r_{oi}) = 0.6 \;\: \mbox{kcal}\cdot \mbox{mol}^{-1} \: .
\end{equation}

To obtain the remaining equation determining the
  parameters $\sigma_i$, $\eps_i$, we require that the DFT transfer free
energy, as computed by the dilute limit of
  equation~(\ref{eq:first}) for the single particle
representing an amino acid side chain, 
should be equal to the experimental value as given in
reference~\cite{AutonM05}, specifically for transfer into a
  solution of 1M TMAO. 
This involves computing the integral over the 
osmolyte-accessible volume in the expression
\begin{equation}
\label{eq:dgint}
\rho \kB T \int d^3 r \left( 1 -\mathrm{e}^{-\beta v_i(r)} \right)
\end{equation}
and setting the result to the empirical value of $\delta g_i$ for
  each amino acid.

The sum of the transfer free energies of each amino acid in a
  Gly-X-Gly tripeptide is often used to approximate the
  conformationally-averaged transfer free energy for a
  protein.\cite{AutonM05} Here we consider the tripeptide transfer
  free energies.
The integral in expression~(\ref{eq:dgint}) then
involves integration over a solid angle $\Omega_i$ determined
by the fraction of solid angle available to the side chain in the
tripeptide {\it vs.} that for the isolated residue, i.e. 
\begin{equation*}
\Omega_i = \frac{A^i_{tri}}{A^i_{iso}}  4\pi
\end{equation*}

The potential $v_i$ is then fully determined from
equation~(\ref{eq:vroi}) along with
\begin{equation}
 \Omega_i \rho \kB T \int_0^\infty dr \: r^2 \left(1 - \mathrm{e}^{-\beta v_i( r )}
\right) = \delta g_i  \: .
\end{equation}

We can now construct potentials for each amino acid transfer free
  energy given in
reference~\cite{AutonM04}. The parameters derived from doing so are listed
in \ref{tab:emp_val}. 
The backbone-osmolyte interaction was parameterized as $v_{BB}(r) =
C/r^{12}$, as this better represented it's strongly repulsive
character. The value of $C$ obtained by fitting to $\delta g_{BB}$ was
$C=7.510\times 10^7$ kcal$\cdot$\AA$^{12}$.  

In this context, the DFT formulation
provides a way of using the information from tri-peptide experiments in
a way that captures both energetic and entropic effects. The
parameters just obtained can be used to determine the change in the transfer free
energies for isolated residues as temperature changes. The
experimental transfer free energies $\delta g_i$ are predicted to
increase as temperature increases, with the new values at $T=310$K given
in \ref{tab:emp_val}. 
Increasing temperature by $0.03\kB T$ increases the transfer free energy
by $\approx 0.6\kB T$ for a 100 residue protein. This change is not
large, but the relative temperature change is also small. The transfer
entropy is significant: $d(\delta g)/dT \approx 20 \kB$.

\begin{table}
\caption{\label{tab:params} 
  Parameter values yielding transfer free energies
    $\delta g$ to 1M TMAO for amino acid side chains and backbone
    at 300K, and the predicted $\delta g$ at 310K .} 
\begin{minipage}{\textwidth}
\begin{tabular}{cccccc}
Type & $ r_o $ (\AA)~\footnote{Distance where the
  osmolyte-amino acid potential is taken to be 0.6
  kcal$\cdot$mol${}^{-1}$} & $\delta g$
(cal/mol)~\footnote{Empirical transfer free energies to 1M
  TMAO} & $\sigma$ (\AA)~\footnote{van der Waals size
  parameter} & $\epsilon$ (kcal/mol)~\footnote{van der
  Waals well depth} & $\delta g (T=310K)$ (cal/mol)~\footnote{
  predicted transfer free energies at $T=310$K} \\
\hline
Ala & 2.52 & -14.64 & 3.517 & 0.6286 & -12.65 \\
\hline
 Arg & 3.28 & -109.3 & 4.088 & 1.022 & -104.0  \\
 \hline
 Asn & 2.74 & 55.69 & 4.564 & 0.0483 & 58.06 \\
 \hline
 Asp & 2.79 & -66.67 & 3.627 & 1.055 & -63.31 \\
 \hline
 Gln & 3.01 & 41.41 & 4.397 & 0.1710 & 44.57 \\
 \hline
 Glu & 2.96 & -83.25 & 3.799 & 0.9973 & -78.88 \\
 \hline
 His & 3.04 & 42.07 & 4.428 & 0.1707 & 45.28 \\
 \hline
 Ile & 3.09 & -25.43 & 4.084 & 0.5692 & -21.59 \\
 \hline
 Leu & 3.09 & 11.6 & 4.246 & 0.3405 & 15.15 \\
 \hline
 Lys & 3.18 & -110.23 & 3.968 & 1.126 & -104.7 \\
 \hline
 Met & 3.09 & -7.65 & 4.154 & 0.4538 & -3.791 \\
 \hline
 Phe & 3.18 & -9.32 & 4.237 & 0.4587 & -5.397 \\
 \hline
 Pro & 2.78 & -137.7 & 3.457 & 1.987 & -133.5 \\
 \hline
 Ser & 2.59 & -39.04 & 3.4905 & 0.8849 & -36.45 \\
 \hline
 Thr & 2.81 & 3.75 & 3.9312 & 0.3889 & 6.41 \\
 \hline
 Trp & 3.39 & -152.9 & 4.157 & 1.150 & -146.5 \\
 \hline
 Tyr & 3.23 & -114.3 & 4.020 & 1.103 & -109.2 \\
 \hline
 Val & 2.93 & -1.02 & 4.021 & 0.4238 & 1.78 \\
 \hline
BB~\footnote{Backbone is parameterized for TMAO by a purely repulsive
  potential (see text)} & 2.25 & 90.0 & - & - & 92.7 \\
\hline
\end{tabular}
\end{minipage}
\label{tab:emp_val}
\end{table}

\section{Using DFT for Implicit Solvent Models}
\label{sec:imp_solv}

The DFT methodology has been applied to the problem of solvation to
calculate fluid correlation functions, solvation free
energies, and reorganization energy in charge
transfer~\cite{RamirezR05,BorgisD12}. 
The use of time-dependent density functional theory has been well-established 
to understand solvation dynamics in single-component solvents~\cite{RoyS93}
as well as selective solvation in binary
mixtures~\cite{ChandraA91,YoshimoriA98}. 
The methodology has also been applied to the connect static and
dynamic approaches to the glass transition by Kirkpatrick and Wolynes~\cite{Kirkpatrick87a}.
The DFT methodology as described above may also be
be applied to the problem of finding the
effective forces for molecular
dynamics simulation in an implicit solvent, which we briefly describe
here.  

We again write the external potential due to solute-solvent
interactions as 
\begin{equation*}
\mathcal{V}(\bm{r}) = \sum_j v_j (|\bm{R}_j - \bm{r}|)
\end{equation*}
we can write the force on the $i$th particle from the transfer free
energy in equation (\ref{eq:large_v}) (neglecting solvent
inter-particle correlations) as 
\begin{align}
\mathbf{F}_i &= \nabla_{R_i} \left[ k_BT\rho \int d^3r \: \left(1 -
    \mathrm{e}^{-\beta \sum_j v_j (|\bm{R}_j - \bm{r}|)}  \right) \right]
\nonumber \\
&= k_BT\rho\beta \int d^3r \: \mathrm{e}^{-\beta \sum_j v_j (|\bm{R}_j
  - \bm{r}|)}  \nabla_{R_i}  v_i (|\bm{R}_i -  \bm{r}|)  \nonumber \\
&= \rho \int d^3r \: \mathrm{e}^{-\beta \sum_j v_j
  (|(\bm{R}_j-\bm{R}_i) - \bm{r}|)} \nabla v_i( r ) 
\label{eq:imp_force}
\end{align}

We immediate see that the integrand is non-zero only when $\nabla v_i(
r )$ is non-zero, so that if there is an effective cutoff $r_c$ such
that $v_i( r ) \approx 0$ for $r > r_c$, then the integral in equation
(\ref{eq:imp_force}) only needs to be taken in the region
$r<r_c$. This is a generalization of the result obtained by
G\"{o}tzelmann {\it et al}\cite{GotzelmannB98}, who have shown that
for a hard sphere potential, only the solvent density at the surface
of the spheres was relevant to the calculation of depletion
forces. Here we extend this analysis to arbitrary potentials.

Consider a particle with a spherically symmetric $v_i( r )$, as
assumed above. The net force on this particle when isolated is
zero. When a second particle exerting potential $v_j(r)$ on the
osmolytes is brought near, the net force on the first due to the
  solvent is a result of
the now asymmetric solvent density. We note here we are treating
  the indirect force rather than the direct force between the
  particles, which can be calculated by direct application of the
  interparticle potential.
 The region of asymmetric solvent density constitutes a
  restricted volume to be integrated over in equation 
(\ref{eq:imp_force}), as only the region of overlap between
the two spheres defined by the cutoff in potential around $\bm{R}_i$
and $\bm{R}_j$ contributes to the net force (see e.g. \ref{fig:fvssep}b below). In addition, the solvent field in
this overlap region will maintain cylindrical symmetry about the axis
joining the two particles, which means that the force will be along
this axis as well. This suggests that the force on particle $i$ can be
written as
\begin{equation*}
\bm{F}_i = \sum_{|\bm{R}_{ij}| < 2r_c} F_{ij} (| \bm{R}_{ij} |)
\hat{\bm{R}}_{ij} \: . 
\end{equation*}
Here $\hat{\bm{R}}_{ij}$ is the unit vector from particle $j$ to
particle $i$, and 
$F_{ij}$ is a scalar function of the interparticle distance 
$| \bm{R}_{ij} | \equiv | \bm{R}_i - \bm{R}_j |$, which is determined by
the overlap integral in equation (\ref{eq:imp_force}), and which could in
principle be pre-computed and tabulated to speed up execution.

\subsection{Depletion and impeded-solvation interactions in an implicit solvent model}
\label{sec:imp_dep}

We can use 
equation~(\ref{eq:imp_force})
to investigate the forces due to solvent on colloidal particles. 
In what follows, we imagine the ``solvent'' to be simplified osmolytes
within an implicit solvent bath. 
This subject has been well-studied (see
e.g. refs.~\cite{OosawaF54,AttardP89,AttardP90,LekkerkerkerHNW92,DickmanR97} );
our goal 
here is simply to show that the DFT transfer free energy provides a
natural way of calculating depletion forces as well as transfer
energies, and that even the approximated form in equation~(\ref{eq:imp_force})
yields non-trivial results for the depletion force. 

We investigate a model consisting of two spheres that interact only by a hard wall potential of radius
$r_s$.
Each sphere also interacts with a bath of osmolytes
through a 6-12 (van der Waals) potential:
$
V( r ) = 4\epsilon \left( \left(\sigma/r\right)^{12} - \left(\sigma/r\right)^6 \right)
$,
with $\sigma = r_s+r_o$.
With this model we examine
the force as a function of the sphere separation $d$. Any force
  between the spheres is entirely due to osmolyte-mediated effects.

\begin{figure}
\includegraphics[width=7.6cm]{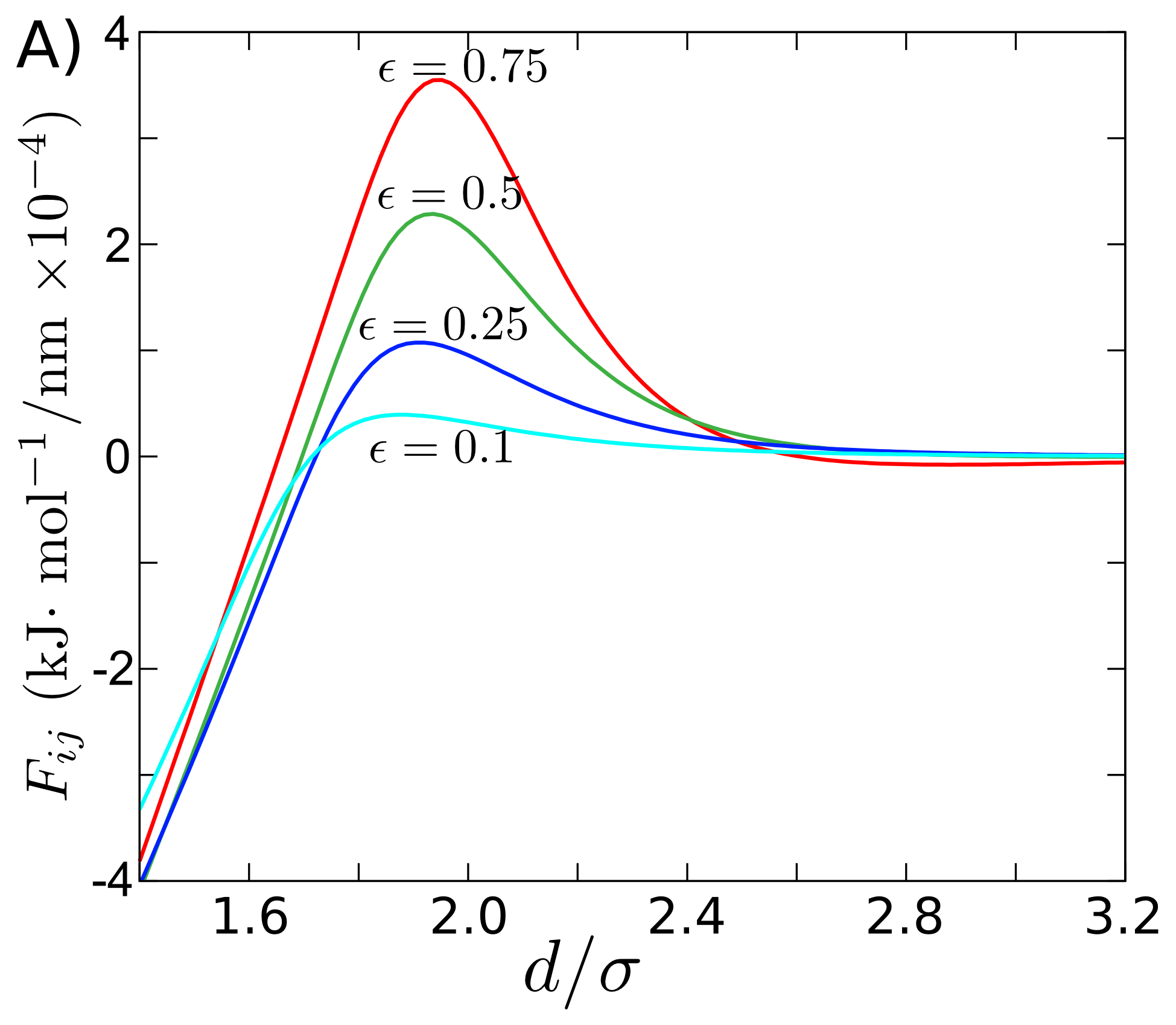}
\includegraphics[width=8.6cm]{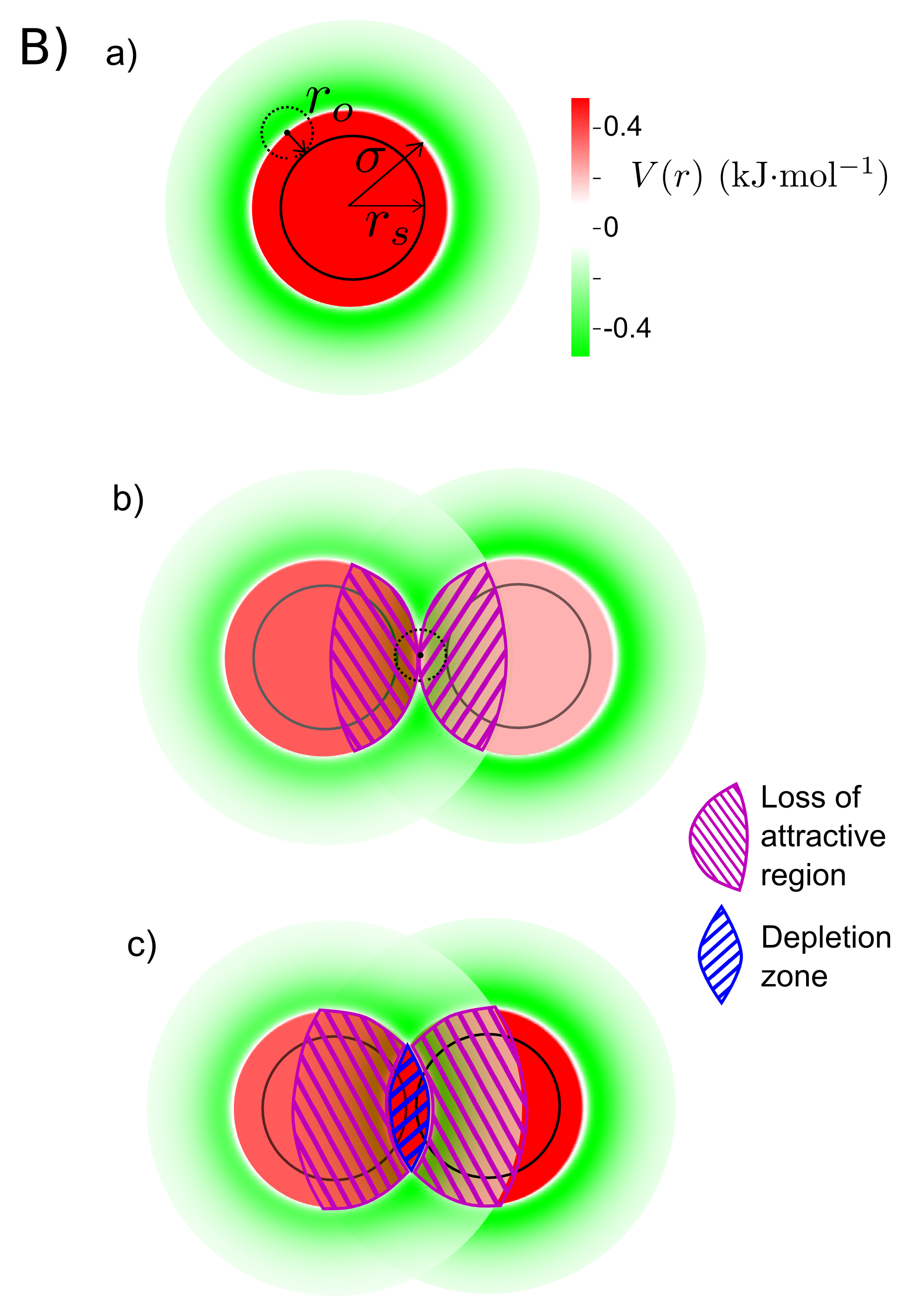}
\caption{
  (A) Solvent-induced force on
  a pair of ``hard-wall'' spheres as a function of the
  separation distance, as obtained from
  equation~(\ref{eq:imp_force}). 
Spheres interact with osmolytes through a LJ potential (see text).
The only parameters that determine the force are thus $\sigma$
  and $\eps$, which appear in the LJ potential that enters into the
  DFT expression for the force.
Each  curve in the figure corresponds to a given well-depth $\eps$ in
  the sphere-osmolyte potential.
The depletion force is dominant at
  small separation, but there is a region in which the spheres are
  mutually repulsive due to lost attraction or ``impeded-solvation'' to the solvent.  
(B) Schematic renderings of the solute spheres in (A) at several
  distances.
  a) The sphere-osmolyte interaction is through a LJ
  potential, which is negative beyond a
  distance $\sigma=r_s+r_o$ (shown as the green region), and positive
    and repulsive
    for $d<\sigma$ (red region). 
The direct sphere-sphere interaction is only through a hard-wall potential
  of radius $r_s$. The osmolytes have radius $r_o$.
  (b) Sphere configuration when
  distance $d=2\sigma$. An osmolyte can just fit between the spheres at
  this distance- the LJ potential is zero in this
  configuration 
if the osmolyte (dashed
    sphere) is centered
    directly between the solute particles.
 Such separations have positive force
between the solutes
in \ref{fig:fvssep}a, due to ``impeded-solvation'':  
 the repulsive interaction between one sphere-osmolyte pair removes some
  of the attractive region from the other sphere-osmolyte pair
  (region shown in magenta). At the separation
  shown in (c), the solvent-induced force between the spheres is now attractive; the volume
  of the removed attractive region now varies
 weakly with separation, and bringing the spheres closer together
  gains free energy by removing the depletion zone highlighted in
  blue.
}
\label{fig:fvssep}
\end{figure}

When the solute particles are far apart, they dress themselves
  with osmolyte solvation shells because of the attractive solute-osmolyte
  potential. As we imagine moving the two solute particles closer
  together, eventually the repulsive region of one solute particle
  overlaps with the attractive region of the other solute particle,
  and vice versa. This situation is unfavorable for the solute
  particles, and the energy may be lowered by moving them further
  apart; hence there is a repulsive force at these distances (see \ref{fig:fvssep}).
  As the
  solute particles continue to approach each other, the above
  repulsive region encroaches on the regions of space where the 
  van der Waals potential is deeper. A larger amount of potentially favorable
  binding energy is removed per distance travelled, and the repulsive
  force due to ``impeded-solvation'' increases. The repulsive force is maximal when the solute
  separation $d$ is roughly $2\sigma$. For separations $d < 2\sigma$,
  the repulsive regions of the two solute spheres begin to overlap. 
This reduces the volume excluded, or more precisely repulsive to,
osmolytes. This reduced excluded volume results in an attractive force
which is the traditional depletion force. Eventually the depletion
force becomes stronger than the above impeded-solvation force, and the
net force becomes attractive. We note that such effects would not be
present in standard GB/SA models of implicit solvation. 

In general, direct inter-particle interactions must be superimposed on the
above scenario.
Which force dominates at a given separation 
will then depend on the values for $r_s$, $r_o$, and $\epsilon$, along with
  the strength of the direct interaction.
The above-described repulsive effect has been observed
before
in hard-sphere solutes using the Derjaguin approximation to obtain an effective surface
tension~\cite{GotzelmannB98}.
Here we see that the effect arises naturally from
the presence of an attractive potential in the density functional theory. 

\section{Conclusions}
\label{sec:conc}

In this paper we have explored the application of the density
functional framework to protein transfer free energies. We have focused primarily on conceptual questions, such
as the role of solvent excluded volume, the temperature dependence of
transfer free energies, and how the density functional theory (DFT)
would reduce to a Volume + SASA model of transfer free energy. 

We compared the DFT results with those from a simplified model that
treated the protein as a tube with a given volume and surface area, on
which osmolytes could condense.
The DFT contains 
contributions from both enthalpy and entropy, so it allows for the calculation of the
temperature-dependence of the transfer free energy.

A further development of the theory presented here which accounts for interparticle
correlations while maintaining computational efficiency is an
important topic for future research. 
As well, the calculation of transfer free energies was implemented here
for a model system with simplified potentials that were parameterized
to experimental values. One could extend this by implementing the theory using
more realistic potential models, and all-atom representations of a protein or
peptide. The various approximations involved in these potentials and models could then be tested and the limits of their validity determined through comparisons with experiment and simulation.
The DFT framework may also provide a method to obtain computationally efficient
but still accurate implicit solvent models for molecular
dynamics simulation, a subject of immense practical importance.
In general, the framework of
density functional theory can provide a powerful tool to explore
aspects of solvation in the context of protein folding, and can do so in a
systematic way. 

\section{Acknowledgements}
S.S.P acknowldeges funding support from PrioNet Canada, NSERC, and
computational support from the WestGrid high-performance computing
consortium. E.A.M. has been supported by the NSERC PGSD program 
for the early part of this work. Lastly, S.S.P. would like to express
his gratitude to Peter G. Wolynes, for providing some of the most
enlightening and enjoyable years of his research career- if only I
knew then what I know now!

\providecommand*{\mcitethebibliography}{\thebibliography}
\csname @ifundefined\endcsname{endmcitethebibliography}
{\let\endmcitethebibliography\endthebibliography}{}

\begin{figure}[p!]
\includegraphics[width=5.08cm]{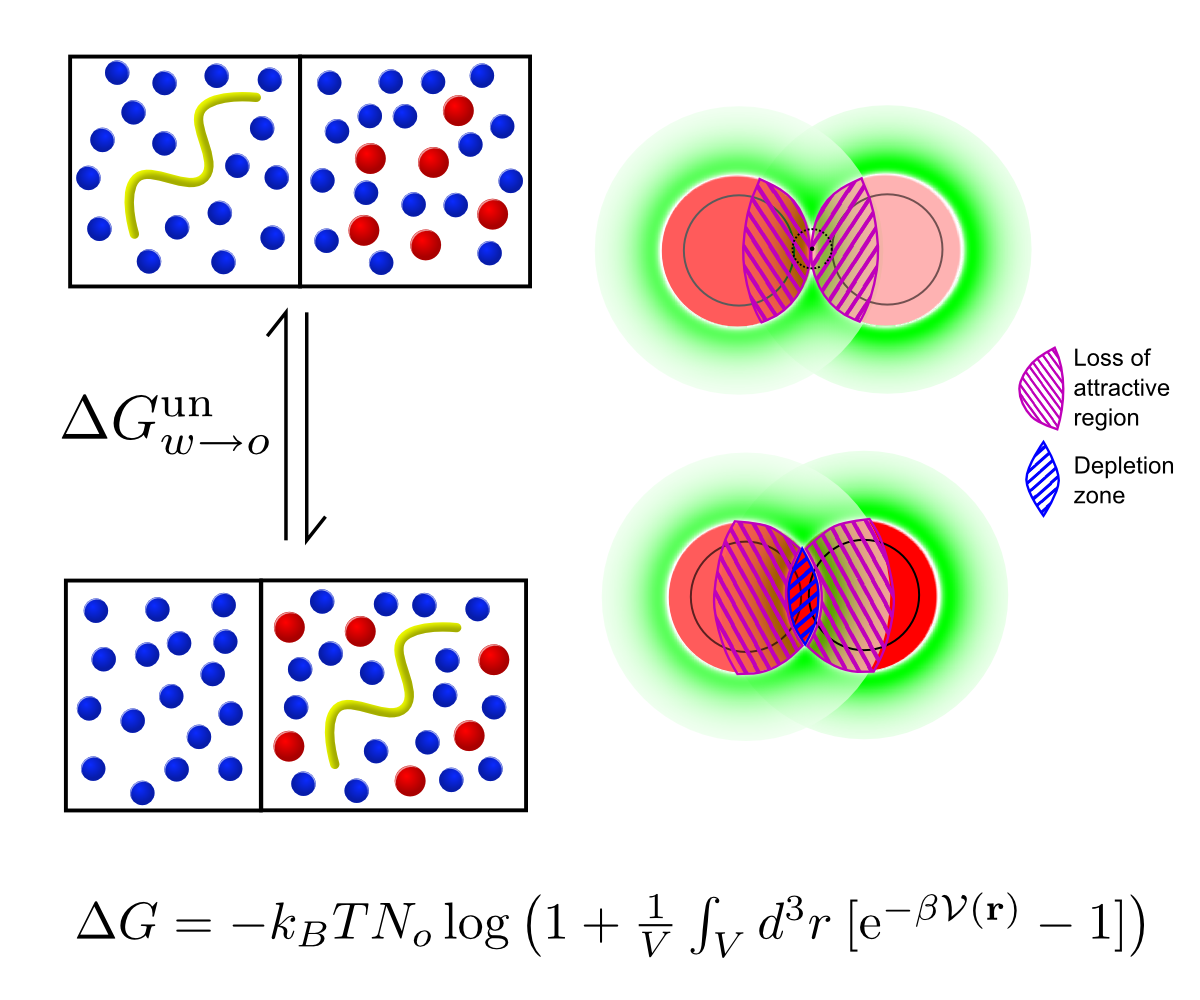}
\end{figure}

\end{document}